\documentclass[reprint,superscriptaddress,amsmath,amssymb,aps,prd,notitlepage,longbibliography,floatfix,nofootinbib,onecolumn]{revtex4-1}

\usepackage{tensor}     
\usepackage{graphicx}   
\usepackage[
colorlinks=true,        
citecolor=blue,         
linkcolor=blue,         
urlcolor=blue           
]{hyperref}             
\usepackage{bm}         
\usepackage{tabulary}
\usepackage{color}      
\usepackage[utf8]{inputenc} 
\usepackage[section]{placeins} 
\usepackage{orcidlink}
\newcommand{\nc}{\newcommand*} 
\nc{\mH}{\mathcal{H}}
\def\[{\left[}
\def\]{\right]}
\def\e{\begin{equation}}
\def\q{\end{equation}}
\def\m{\begin{eqnarray}}
\def\n{\end{eqnarray}}

\nc{\Eq}[1]{Eq.~\eqref{#1}}     
\nc{\Fig}[1]{Fig.~\ref{#1}}     
\nc{\Table}[1]{Table~\ref{#1}}  
\nc{\Sec}[1]{Sec.~\ref{#1}}     
\nc{\fpbh}{f_{\mathrm{pbh}}}
\nc{\ogw}{\Omega_{\mathrm{GW}}}
\nc{\eg}{\textit{e.g.~}}
\nc{\app}{\approx}
\nc{\hf}{\frac{1}{2}}
\nc{\red}[1]{\textcolor{red}{#1}}
\nc{\yc}[1]{{\textcolor{red}{\sf{[YC: #1]}} }}
\begin{document}
\title{GW231123 Mass Gap Event and the Primordial Black Hole Scenario} 

\author{Chen Yuan\orcidlink{0000-0001-8560-5487}}
\affiliation{CENTRA, Departamento de Física, Instituto Superior Técnico – IST, Universidade de Lisboa – UL, Avenida Rovisco Pais 1, 1049–001 Lisboa, Portugal}
\author{Zu-Cheng Chen\orcidlink{0000-0001-7016-9934}}
\email{zuchengchen@hunnu.edu.cn}
\affiliation{Department of Physics and Synergetic Innovation Center for Quantum Effects and Applications, Hunan Normal University, Changsha, Hunan 410081, China}
\affiliation{Institute of Interdisciplinary Studies, Hunan Normal University, Changsha, Hunan 410081, China}
\author{Lang~Liu\orcidlink{0000-0002-0297-9633}}
\email{liulang@bnu.edu.cn}	
\affiliation{Faculty of Arts and Sciences, Beijing Normal University, Zhuhai 519087, China}
	
\begin{abstract}
We investigate the possibility that the recently reported GW231123 event, with component masses $M_1=137^{+22}_{-17}\,M_\odot$, $M_2=103^{+20}_{-52}\,M_\odot$ and a local merger rate $R_{\mathrm{local}}=0.08^{+0.19}_{-0.07}\,\mathrm{Gpc^{-3}\,yr^{-1}}$, originates from primordial black holes (PBHs) formed during an early matter-dominated era. We compute the PBH mass function, abundance, spin distribution and the merger rate density and find a set of choices for the parameters to reproduce the key properties of GW231123. While PBHs formed in such a scenario can acquire large spins through sustained tidal torques, the spin distribution remains uncertain and additional accretion might lead to extreme spin values inferred in GW231123. We also show that the resulting PBH abundance, $f_{\mathrm{pbh}}=1.64^{+5.00}_{-1.59}\times10^{-1}$, lies close to the exclusion bounds from CMB accretion limits and other probes, highlighting a potential tension with current constraints. Finally, we estimate the scalar-induced gravitational waves (SIGWs) that are inevitably generated during PBH formation. PBHs that interpret GW231123 are accompanied by negligible SIGWs in the nano-hertz band, indicating no conflict with current pulsar timing arrays data.

\end{abstract}

\maketitle
\section{Introduction}

The era of gravitational wave (GW) astronomy began with the landmark detection of GW150914 by the LIGO-Virgo collaboration~\cite{LIGOScientific:2016emj}. Subsequent observing runs expanded the catalog to nearly a hundred GW events~\cite{LIGOScientific:2017vwq,LIGOScientific:2016sjg,LIGOScientific:2018mvr,LIGOScientific:2020ibl,KAGRA:2021vkt}, providing a unique way to test gravity in the strong-field regime. The ongoing fourth observing run has recently reported GW231123~\cite{LIGOScientific:2025rsn}, an extraordinary event with component masses $M_1 = 137^{+22}_{-17}\,M_\odot$ and $M_2 = 103^{+20}_{-52}\,M_\odot$. These masses place both components within or above the theorized pair-instability mass gap $\sim [60,130]\,M_\odot$~\cite{Woosley:2016hmi,Farmer:2019jed,Woosley:2021xba,Hendriks:2023yrw}, directly challenging standard stellar-collapse formation channels for black holes.

The pair-instability mechanism predicts that stars with helium cores in the range $\sim 32$--$64\,M_\odot$ undergo pulsational pair-instability, ejecting sufficient mass to produce black holes below $\sim 60\,M_\odot$. More massive cores experience complete disruption through pair-instability supernovae, leaving no remnant and creating an upper bound at $\sim 130\,M_\odot$ where direct collapse resumes. While astrophysical scenarios such as hierarchical mergers or stellar collisions have been proposed to populate this forbidden region \cite{Kimball:2020qyd,Mahapatra:2021hme, Mould:2022ccw, Wang:2022gnx, Li:2023yyt, Pierra:2024fbl, Hussain:2024qzl, Antonini:2024het}, GW231123 presents an additional challenge: both components exhibit exceptionally high dimensionless spins of $\chi_1 = 0.90^{+0.10}_{-0.19}$ and $\chi_2 = 0.80^{+0.20}_{-0.51}$. Producing such massive black holes with near-maximal spins through conventional astrophysical channels requires extreme fine-tuning of multiple rare processes.

A compelling alternative to astrophysical black holes is primordial black holes (PBHs), which form in the early Universe via the gravitational collapse of large curvature perturbations~\cite{Zeldovich:1967lct,Hawking:1971ei,Carr:1974nx}. PBHs can span a broad mass range, potentially accounting for some or all of the dark matter~\cite{Khlopov:2008qy,Sasaki:2018dmp,Carr:2020gox,Carr:2020xqk,Green:2020jor,LISACosmologyWorkingGroup:2023njw,Byrnes:2025tji}, while also providing explanations for GW events detected by the LIGO-Virgo-KAGRA (LVK) collaborations if even a small fraction of dark matter consists of PBHs~\cite{Sasaki:2016jop,Clesse:2016vqa,Chen:2018czv,Raidal:2018bbj,Liu:2018ess,Liu:2019rnx,Carr:2019kxo,DeLuca:2020qqa,Hall:2020daa,Liu:2022iuf,Bhagwat:2020bzh,Hutsi:2020sol,Clesse:2020ghq,Wong:2020yig,DeLuca:2021wjr,Mukherjee:2021ags,Bavera:2021wmw,Franciolini:2021tla,Chen:2021nxo,Chen:2024dxh,Huang:2024wse,Yuan:2024yyo,Afroz:2024fzp,Yogesh:2025hll,Crescimbeni:2025ywm}.

The spin properties of GW231123 suggest a crucial discriminant between formation scenarios. In the standard radiation-dominated collapse scenario, PBH spins originate from first-order tidal gravitational fields generating torques when the corresponding wavelength re-enters the horizon. This first-order effect yields PBHs with near-zero initial spins~\cite{Mirbabayi:2019uph,DeLuca:2019buf,Banerjee:2024nkv,Harada:2020pzb}, inconsistent with the high spins inferred in GW231123 \footnote{The accretion processes can potentially spin up PBHs over cosmic time, achieving the extreme spins $\chi$ around $0.8-0.9$ observed in GW231123~\cite{DeLuca:2025fln}.}. However, during a transient matter-dominated (MD) phase, the physics changes dramatically. Growing density perturbations lead to sustained tidal torques that continuously spin up collapsing regions, potentially reaching near-maximal values. Consequently, PBHs generated during an MD era can naturally acquire large spins~\cite{Harada:2017fjm} \footnote{The spin distribution of PBHs formed during the early matter-dominated era is uncertain. There are differing opinions in the literature, with Ref. \cite{Ye:2025wif} suggesting that their spin distribution might be small.}, offering a natural explanation for the extreme properties of GW231123.

On the other hand, the enhanced linear curvature perturbations required to produce an appreciable PBH abundance inevitably couple at second order, generating scalar-induced gravitational waves (SIGWs)~\cite{Yuan:2021qgz,Domenech:2021ztg}. These SIGWs provide an independent observational probe of the PBH formation scenario, potentially detectable by pulsar timing arrays (PTAs) in the nano-hertz band. The SIGWs thus provide another independent test of the PBH interpretation.

In this letter, we investigate whether GW231123 can be explained by PBHs formed during an early MD era. We compute the PBH mass function, spin distribution, and merger rate density, searching for appropriate parameter choices that can reproduce all key properties of GW231123. Then we investigate the required PBH abundance in dark matter to interpret GW231123 and compare it with current observational constraints. Furthermore, we calculate the associated SIGWs and discuss their detectability by PTAs.

\section{PBH formation during the MD era and SIGW}\label{MDpbh}
In this work we focus on the formation of PBHs during an early MD era, characterized by an equation of state $w=0$. 
We adopt a widely used log‑normal power spectrum \cite{Yuan:2019wwo,Pi:2020otn,Domenech:2020kqm,Yuan:2020iwf,Yuan:2021qgz,Meng:2022ixx,Yuan:2023ofl,Gorji:2023ziy,Franciolini:2023pbf,Papanikolaou:2024kjb} to characterize the enhancement of  curvature perturbations on small scale, namely
\begin{equation}\label{Pzeta}
    \mathcal P_\zeta(k)
    = \frac{A}{\sqrt{2\pi}\Delta} \exp\biggl[-\frac{\ln^2(k/k_*)}{2\,\Delta^2}\biggr],
\end{equation}
where $A$ and $\Delta$ is related to the amplitude and the width of the spectrum respectively and $k_*$ represents the peak wavelength of the enhancement. We focus on the narrow spectrum case where $\Delta \lesssim 0.1$ so that the PBH mass function is concentrated at a certain value, just enough to explain GW231123 while suppressing any overproduction of lower‐mass binaries in the O3 catalog.
An important quantity related to PBH formation is the smoothed variance of the power spectrum
\begin{equation}\label{variance}
    \sigma_H^2(M)
    =\Bigl(\frac{2+2w}{5+3w}\Bigr)^{2}
    \int_0^\infty \frac{dq}{q}\,\left(\frac{q}{k}\right)^4\,\mathcal P_\zeta(q)\,W^2(q/k),
\end{equation}
where $W(x) = \exp(-x^2/4)$ is the Gaussian window function.
In a general $w$-dominated Universe, the masses of PBHs are connected to the comoving wavelength, $k$, by \cite{Bhattacharya:2019bvk,Liu:2023pau,Liu:2023hpw}
\begin{align}\label{mpbh}
\frac{M}{M_\odot}&\approx  0.01\frac{\bar{\gamma}}{0.2}\!\left(\frac{k_{\rm rh}}{k}\right)^{\!\frac{3(1+w)}{1+3w}}\! \left(\frac{106.75}{g_{*r}(T_{\rm rh})}\right)^{\frac12}\!\left(\frac{{\rm GeV}}{T_{\rm rh}}\right)^{\! 2},
\end{align}
where $\bar{\gamma}\approx0.2$ represents the fraction of matter within the Hubble horizon that undergoes collapse to form PBHs. Here $k_{\rm rh}$ and $T_{\rm rh}$ are the comoving wavelength and temperature of the Universe at reheating time respectively. The temperature $T$ at which a mode with wavenumber $k$ re-enters the horizon can be estimated as 
\begin{equation}
\label{k-T}
    k\! \simeq \frac{\!1.5\!\times\!  10^7}{\mathrm{Mpc}}
    \left({g_{*r}(T)} \over {106.75} \right)^{1\over 2}
    \left({g_{*s}(T)} \over 106.75\right)^{-\frac{1}{3}}
    \left (\frac{T}{\rm GeV}\right).
\end{equation}

The probability of PBH formation during the MD era is studied in \cite{Harada:2016mhb} and the results are well fitted by $\beta \simeq 0.05556\sigma_H^5(M)$. One can define the fraction of PBHs in the cold dark matter at present as
\begin{equation}\label{fpbh}
f(M) \equiv\frac{\Omega_\text{PBH}(M)}{\Omega_\text{CDM}} \approx 1.5\times 10^{13}\beta\left(\frac{k}{k_{\rm rh}}\right)^{\frac{6w}{1+3w}}\left(\frac{T_{\rm rh}}{{\rm GeV}}\right) \left(\frac{g_{*s}(T_{\rm rh})}{106.75}\right)^{-1}\left(\frac{g_{*r}(T_{\rm rh})}{106.75}\right), 
\end{equation}
where $g_{*r}$ and $g_{*s}$ are the effective energy and entropy degrees of freedom, respectively. 
As a result, one can express the PBH mass function as
\begin{equation}
P(M) \equiv \frac{f(M)}{\fpbh M},
\end{equation}
where $\fpbh \equiv \int f(M) \mathrm{d}\ln M $ is total abundance of PBH in cold dark matter. One can readily verify that $\int P(M)\mathrm{d}M = 1$ in this convention.

During the formation of PBHs, the first-order scalar modes are coupled to induce tensor perturbations at second-order, known as the SIGWs. We compute SIGWs in a $w$-dominated Universe based on the analytical results from \cite{Domenech:2020kqm,Domenech:2021ztg}. We assume that the Universe reheats as soon as the $k_\mathrm{rh}$-mode re-enters the horizon. The dimensionless energy spectrum of SIGWs is given by
\begin{equation}\label{ogw0}
\Omega_{\mathrm{GW}, 0} h^2 \approx  1.62 \times 10^{-5} \left(\frac{g_{*r}\left(T_{\mathrm{rh}}\right)}{106.75}\right)\left(\frac{g_{* s}\left(T_{\mathrm{rh}}\right)}{106.75}\right)^{-\frac{4}{3}} \left(\frac{\Omega_{r, 0} h^2}{4.18 \times 10^{-5}}\right)  \Omega_{\mathrm{GW}, \mathrm{rh}}, 
\end{equation}
where $\Omega_{r,0}h^2 \approx 4.18 \times 10^{-5}$ denotes the current radiation density parameter \cite{Planck:2018vyg}. For modes with $k \gtrsim k_{\rm rh}$, the SIGW spectrum in Eq.~\eqref{ogw0} takes the form \cite{Domenech:2020kqm}
\begin{equation}\label{ogwrh}
\Omega_{\mathrm{GW},\mathrm{rh}}=\left(\frac{k}{k_{\mathrm{rh}}}\right)^{\! -2 b} \int_0^{\infty} d v \int_{|1-v|}^{1+v} d u\, \mathcal{T}(u,v,w)\, \mathcal{P}_{\mathcal{R}}(k u) \mathcal{P}_{\mathcal{R}}(k v),
\end{equation}
where $b\equiv (1-3w)/(1+3w)$, and the explicit form of $\mathcal{T}$ is provided in~\cite{Domenech:2020kqm,Domenech:2021ztg}. For $k \lesssim k_{\rm rh}$, we find $\Omega_{\mathrm{GW},\mathrm{rh}} \propto (k/k_{\rm rh})^2$, which suppresses the SIGW spectrum amplitude by $(k_{\rm rh}/k)^2$. This suppression arises from the following mechanism: before the scalar mode with wavenumber $k$ enters the horizon, tensor modes with wavenumber $k$ experience a constant source term, causing their amplitude to grow as $(k\tau)^2$, where $\tau$ denotes conformal time. Once the $k$ scalar mode reenters the horizon around $\tau \sim1/k$, the source term vanishes and tensor mode growth stops. Therefore, maintaining a fixed SIGW spectrum amplitude at the reheating wavenumber requires larger scalar perturbation amplitudes as the peak scale moves farther from the reheating scale.

The present-day GW frequency $f$ corresponding to wavenumber $k$ is 
\begin{equation}
\label{k-f}
    f= \frac{k}{2 \pi} \simeq 1.6\, {\rm nHz}
    \left( \frac{k}{10^6\,{\rm Mpc}^{-1}} \right).
\end{equation}
Combining Eqs. \eqref{k-T} and \eqref{k-f} yields the relationship between frequency $f$ and temperature $T$:
\begin{equation}
    f \simeq 24\, {\rm nHz} \left({g_{*r}(T)} \over {106.75} \right)^{1\over 2}
    \left({g_{*s}(T)} \over 106.75\right)^{-\frac{1}{3}}
    \left (\frac{T}{\rm GeV}\right).
\end{equation}

The requirement for successful BBN sets a lower limit on the reheating temperature of $T_{\mathrm{rh}}\geq 4{\rm MeV}$~\cite{Kawasaki:1999na,Kawasaki:2000en,Hannestad:2004px,Hasegawa:2019jsa}. This translates to a constraint on the reheating frequency $f_{\mathrm{rh}}$ associated with the reheating mode $k_{\mathrm{rh}}$, requiring $f_{\mathrm{rh}}\gtrsim 0.1 \, {\rm nHz}$.

\section{PBH Scenario for GW231123}\label{merge}
We now confront the GW231123 event with a PBH‑binary merger hypothesis. GW231123 is characterized by component masses
$M_1 = 137^{+22}_{-17}\,M_\odot$ and $ M_2 = 103^{+20}_{-52}\,M_\odot$ and an inferred merger rate
$  R_{\rm local} = 0.08^{+0.19}_{-0.07}\,\mathrm{Gpc^{-3}\,yr^{-1}}$. 
Following \cite{Raidal:2018bbj}, the PBH merger rate density for a general PBH mass function is given by
\begin{equation}\label{R12}
\begin{aligned}
&R_{12}(z, m_1, m_2) = \frac{1.6 \times 10^6}{\mathrm{Gpc}^3 \, \mathrm{yr}} \,
f_{\mathrm{pbh}}^{\frac{53}{37}} \left( \frac{t(z)}{t_0} \right)^{-\frac{34}{37}} 
\eta^{-\frac{34}{37}} \left( \frac{M}{M_\odot} \right)^{-\frac{32}{37}} S\left[M, f_{\mathrm{pbh}}, P(m), z\right] 
P(m_1) P(m_2),
\end{aligned}
\end{equation}
where $M = m_1 + m_2$ is the total mass, $\eta = m_1 m_2 / M^2$ is the symmetric mass ratio, $t(z),t_0$ are the age of the Universe at redshift $z$ and today respectively. The S-factor depends on the PBH mass function and its explicit expression can be found in \cite{Hutsi:2020sol,Franciolini:2022tfm}. We also include the three body effect, whose merger rate density reads \cite{Vaskonen:2019jpv,Raidal:2024bmm}
\begin{equation}
    \begin{aligned}
R_{12,3}(z, m_1, m_2)& \approx \frac{7.9 \times 10^4}{\mathrm{Gpc}^3 \, \mathrm{yr}} \left( \frac{t}{t_0} \right)^{\frac{\gamma}{7} - 1} f_{\mathrm{pbh}}^{\frac{144 \gamma}{259} + \frac{47}{37}}  \left[ \frac{\langle m \rangle}{M_\odot} \right]^{\frac{5 \gamma - 32}{37}} \left( \frac{M}{2 \langle m \rangle} \right)^{\frac{179 \gamma}{259} - \frac{2122}{333}} (4 \eta)^{-\frac{3 \gamma}{7} - 1} \\
&\times \mathcal{K} \frac{e^{-3.2(\gamma - 1)} \gamma}{28/9 - \gamma} \, \overline{\mathcal{F}}(m_1, m_2) \, P(m_1) \, P(m_2),
\end{aligned}
\end{equation}
where $\gamma=1$ and $\mathcal{K}=4$ represent the angular momentum distribution and binary hardening. The average PBH mass is evaluated as $\left\langle m \right\rangle = \int M P(M)\mathrm{d}M$ and the function $\overline{\mathcal{F}}$ can be found in \cite{Raidal:2018bbj}.
Rather than scanning the full parameter space, it is instructive to make a simple estimate to fix the range of parameters. 
For a sufficiently narrow power spectrum, Eq.~(\ref{Pzeta}) may be approximated by a Dirac delta function. It then follows from Eq.~(\ref{variance}) that the variance is approximately
\begin{equation}
    \sigma_H(M)
    \simeq\sqrt{A}\Bigl(\frac{2+2w}{5+3w}\Bigr)
    \left(\frac{k_*}{k}\right)^2 \exp\left(-\frac{(k_*/k)^2}{4}\right)
\end{equation}
Neglecting the contribution from the relativistic degrees of freedom and substituting this result into Eq.~(\ref{fpbh}), one can obtain the peak mass, $M_{\rm peak}$, at which $f(M)$ attains its maximum
\begin{equation}
    M_{\mathrm{peak}} \simeq 0.8 M_{\odot}\frac{T_{\rm rh}/{\rm GeV}}{(T_* / {\rm GeV}  )^3}.
\end{equation}
Here, $T_*$ represents the temperature of the Universe when $k_*$ re-enters the horizon. For PBHs formed during MD before the reheating stage, one necessarily has $T_* > T_{\rm rh}$, which lead to
\begin{equation}
M_{\mathrm{peak}} < 0.8 M_{\odot}\frac{1}{(T_* / {\rm GeV}  )^2}.    
\end{equation}
If the two mass components of GW231123 have primordial origin, we should have $M_{\mathrm{peak}}\gtrsim 100 M_{\odot}$, which in turn translates into the bound
\begin{equation}
    T_{\rm rh}\lesssim 0.09 \mathrm{GeV}.  
\end{equation}
Hence, interpreting GW231123 in terms of PBHs formed during the MD era effectively places an upper limit on the reheating temperature.

\begin{figure*}
	\centering
	\includegraphics[width = 0.45\textwidth]{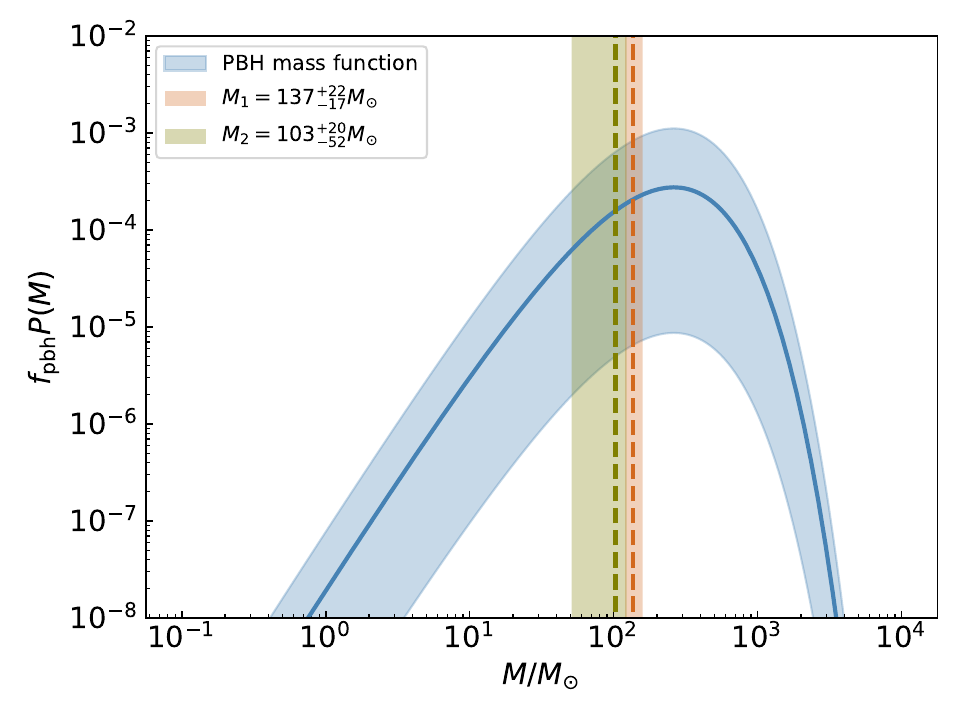}
    \includegraphics[width = 0.45\textwidth]{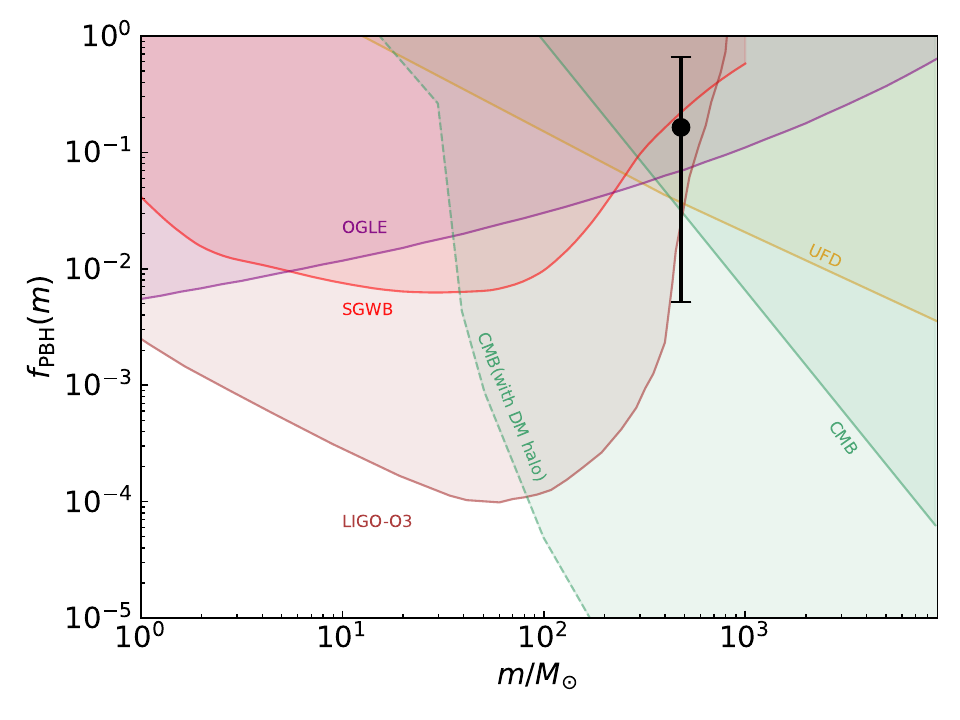}
    \caption{\label{Pf} Left panel: The PBH mass function assuming GW231123 has primordial origin. Right panel: The abundance of PBHs for interpreting GW231123 together with constraints from OGLE microlensing \cite{Mroz:2024mse,Mroz:2024wag}, the null detection of SGWB of binary PBHs \cite{Nitz:2022ltl,Boybeyi:2024mhp}, the CMB accretion results \cite{Ali-Haimoud:2016mbv,Blum:2016cjs,Horowitz:2016lib,Chen:2016pud,Poulin:2017bwe} ({showing both the conservative CMB accretion limit from \cite{Ali-Haimoud:2016mbv} and the more stringent spherical–halo bounds from \cite{Serpico:2020ehh}}), ultra-faint dwarf (UFD) galaxies \cite{Brandt:2016aco} and the LIGO-O3 results assuming all the O3 events have astrophysical origin \cite{Hutsi:2020sol,Andres-Carcasona:2024wqk}.}
\end{figure*}

We fix $\Delta=0.01$, $T_{\rm rh} = 20$MeV and we find a value $k_*=2.3\times 10^{5}~\mathrm{Mpc}^{-1}$ for which the corresponding PBH mass function is consistent with the two mass components of GW231123, as shown in the left panel of Fig.~\ref{Pf}. To fit the local merger rate of GW231123, we obtain 
\begin{equation}\label{Afpbh}
    A = 7.64^{+5.72}_{-5.72}\times10^{-6}, \qquad \fpbh=  1.64^{+5.00}_{-1.59}\times10^{-1}.
\end{equation}

Fig.~\ref{Pf} demonstrates the PBH mass function and the resulting abundance together with current constraints.
If GW231123 is interpreted as PBHs formed during the MD era, we show that the inferred PBH abundance could lie essentially at the interface of the LIGO-O3 and the stochastic GW background (SGWB) excluded regions. The LIGO-O3 limit assumes that all O3 events originate from astrophysical channels \cite{Hutsi:2020sol,Andres-Carcasona:2024wqk}, whereas the SGWB results are derived from the null detection of a stochastic GW background from binary PBHs in the first three runs \cite{Nitz:2022ltl,Boybeyi:2024mhp}. Furthermore, microlensing constraints from OGLE and limits from accreting PBHs by CMB partially overlap with our result.
Thus, the PBH scenario can reproduce the properties of GW231123, but the required abundance is {consistent only with the conservative CMB limits, and strongly disfavored once the spherical–halo accretion bounds are adopted. This illustrates the large theoretical uncertainty associated with PBH accretion modeling.}

On the other hand, the spin of the two mass components of GW231123 is reported to be $0.9^{+0.10}_{-0.19} $ and $ 0.80^{+0.20}_{-0.51}$. Given that PBHs formed in an MD era can acquire large angular momentum through tidal torques, we also compute the spin distribution based on \cite{Harada:2017fjm}:
\begin{equation}
f_{\mathrm{BH}(1)}(a_*)\,\mathrm{d}a_*
\propto
\frac{1}{a_*^3}
\exp\!\Biggl(
  -\frac{1}{2\sigma_H^2}
  \,\frac{3^2\,2^4\,q^4}{5^6\,a_*^4}
\Biggr)
\,\mathrm{d}a_*,
\end{equation}
where $q=\sqrt{2}$ is a parameter related to the initial quadrupole moment of the mass. We adopt the spin distribution from Ref.~\cite{Harada:2017fjm}, while acknowledging that this remains uncertain. We use this distribution as an illustrative case to explore whether MD-era formation could in principle explain GW231123, recognizing that more detailed calculations are needed. For the parameters in Eq.~(\ref{Afpbh}), we found that the above results give an extremely narrow distribution at $a_*=1$, indicating PBHs interpreting GW231123 would have extreme spin. This is consistent with the spin estimate of GW231123.

Finally, we evaluate the SIGWs based on the result Eq.~(\ref{Afpbh}). In this paper, we assume a sudden MD-RD transition where the transition time is much shorter than the Hubble time. In such a case, so that the log-normal spectrum with $\Delta=0.01$ is a good approximation and the energy spectrum of the SIGW generated by such a narrow power spectrum can be evaluated as \cite{Kohri:2018awv}
\begin{equation}\label{SIGW}
\Omega_{\rm GW,0}(k) \;\simeq\; 
\Omega_r\frac{3}{25}\left(\frac{k_\ast}{k_{\rm rh}}\right)^{2}
\left(1 - \left(\frac{k}{2k_\ast}\right)^{2}\right)^{2}
A_\zeta^{2}\,\Theta(2k_\ast - k)\,
\mathcal{R}\!\left(\frac{k}{k_{\rm rh}}\right),
\end{equation}
where $\Omega_r$ is the the energy density parameter for radiation and the suppression factor is given by \cite{Kohri:2018awv}
\begin{equation}
\mathcal{R}(x) \;=\; 
\frac{1}{4x^{8}}
\Bigg[
\Big(-6x + 4x^{3}\cos x + 2x^{4}\sin x + 6\cos x \sin x\Big)^{2}
+
\Big(-3 + 6x^{2} + 2x^{4}\cos x + 3\cos(2x) - 4x^{3}\sin x\Big)^{2}
\Bigg].
\end{equation}
The factor $\mathcal{R}$ originates from solving the equation of motion for the SIGWs across the MD$\to$RD transition: one performs the time integral of the MD scalar source with the Green's function and then matches the tensor mode and its first derivative continuously to the RD solution at the reheating time. Physically, $\mathcal{R}$ quantifies how much of the MD–generated SIGWs is transmitted into the RD era.

According to Eq.~(\ref{SIGW}), the energy spectrum of SIGWs by today is then $\Omega_{\rm GW,0}^{\rm max} \sim \mathcal{O}(10^{-16})$. Although the peak frequency of SIGWs is $f_{*}\simeq 0.4$nHz according to Eq.~(\ref{k-f}), close to the PTA band, the amplitude of $\Omega_{\rm GW,0}$ is far below the sensitivity of PTAs.


\section{Summary and discussion}\label{summary}

In this work, we have demonstrated that a population of PBHs formed during the MD era, seeded by a log-normal curvature power spectrum, can provide a potential explanation for the key properties of GW231123 including the two  mass components $M_1=137^{+22}_{-17}M_\odot$, $M_2=103^{+20}_{-52}M_\odot$, the exceptionally high spins $\chi_1 = 0.90^{+0.10}_{-0.19}$, $\chi_2 = 0.80^{+0.20}_{-0.51}$, and the local merger rate $R_{\mathrm{local}}=0.08^{+0.19}_{-0.07}\,\mathrm{Gpc^{-3}\,yr^{-1}}$.

In our analysis we choose a reheating temperature 
$T_{\rm rh} = 20~{\rm MeV}$, which safely exceeds the BBN requirement $T_{\rm rh} \gtrsim 4~{\rm MeV}$ and fixes the reheating mode at 
$k_{\rm rh} \simeq 3 \times 10^{5}~{\rm Mpc}^{-1}$. 
We further place the curvature power spectrum peak close to reheating, with $k_\ast = 2.3 \times 10^{5}~{\rm Mpc}^{-1}$ 
and a narrow width $\Delta = 0.01$. 
Such a choice spectrum ensures that the MD phase relevant for amplified modes is narrow. The resulting SIGW spectrum peaks at 
$
\Omega_{\rm GW,0}\, h^2 \;\lesssim\; \mathcal{O}(10^{-17})$, which is orders of magnitude below the value that would affect $\Delta N_{\rm eff}$ at BBN or CMB epochs $\Omega_{\mathrm{GW}}\gtrsim \mathcal{O}(10^{-6})$ \cite{Kohri:2018awv}. 
In other words, the combination of a high enough reheating temperature, a narrow peak, and negligible SIGWs keeps the MD era short and the total GW energy density safely within $\Delta N_{\rm eff}$ bounds.

The required PBH abundance for interpreting GW231123 is $f_{\mathrm{pbh}}= 1.64^{+5.00}_{-1.59}\times10^{-1}$, which remains self-consistent with existing observational constraints, though it partially overlaps with current exclusion regions. The MD formation scenario can also account for the high spins observed in GW231123, as sustained tidal torques continuously spin up the collapsing regions. On the other hand, our inferred PBH mass function has an average value of about $400M_{\odot}$. This indicates that if GW231123 indeed originates from PBHs formed during the MD era, we might also expect other GW231123-like events to appear in the near future.

Moreover, our calculated SIGW spectrum falls well below current PTA limits. This demonstrates that the enhanced primordial perturbations needed to generate sufficient PBHs during the MD era do not conflict with current PTA results, leading to a consistency check for the PBH scenario.

Our results depend moderately on the collapse efficiency parameter $\bar{\gamma}$. While we used the standard value $\bar{\gamma} = 0.2$, variations within the range $0.05-0.6$ suggested by recent works can be accommodated by adjusting the power spectrum parameters \cite{Harada:2016mhb, Musco:2018rwt,Musco:2020jjb}. This introduces an additional systematic uncertainty of roughly a factor of 2 in the choice of $k_*$, but does not qualitatively change our conclusions about the viability of the PBH interpretation for GW231123.

Finally, we expect several future observations to aid  in testing the PBH interpretation of GW231123. The forthcoming O4/O5 search for a SGWB from binary PBH mergers could extend the current exclusion regions. Meanwhile, increasingly sensitive microlensing surveys will further constrain the abundance of PBHs, potentially confirming or excluding the PBH scenario for GW231123-like events \cite{Mroz:2025xbl}.

In conclusion, our analysis shows that GW231123 could in principle be interpreted as a PBH merger formed in an early MD era, {but the viability of this explanation crucially depends on the choice of CMB accretion limits: it is marginally allowed under conservative assumptions, yet in strong tension with more stringent spherical–halo bounds. Future improvements in modeling PBH accretion are therefore essential to clarify this possibility.}

\begin{acknowledgments}
C.Y. acknowledge the financial support provided under the European Union’s H2020 ERC Advanced Grant “Black holes: gravitational engines of discovery” grant agreement no. Gravitas–101052587. Views and opinions expressed are however those of the author only and do not necessarily reflect those of the European Union or the European Research Council. Neither the European Union nor the granting authority can be held responsible for them. We acknowledge support from the Villum Investigator program supported by the VILLUM Foundation (grant no. VIL37766) and the DNRF Chair program (grant no. DNRF162) by the Danish National Research Foundation.
ZCC is supported by the National Natural Science Foundation of China under Grant No.~12405056, the Natural Science Foundation of Hunan Province under Grant No.~2025JJ40006, and the Innovative Research Group of Hunan Province under Grant No.~2024JJ1006. 
LL is supported by the National Natural Science Foundation of China (Grant No.~12505054 and ~12433001).
\end{acknowledgments}

\bibliography{ref}
\end{document}